\documentclass[conference]{IEEEtran}
\usepackage{verbatim}
\usepackage{longtable}
\usepackage{xcolor}
\usepackage{amsmath}
\usepackage[ruled,vlined,linesnumbered]{algorithm2e}
\usepackage{stfloats}
\usepackage{etoolbox}
\usepackage{graphicx}
\BeforeBeginEnvironment{figure}{\vskip-2ex}
\AfterEndEnvironment{figure}{\vskip-1ex}
\usepackage{float}
\usepackage{algorithmicx}
\usepackage{fancyhdr}
\usepackage{cite}
\usepackage{booktabs}

\usepackage{amsmath}
\usepackage{longtable}

\usepackage{dirtytalk}
\ifCLASSOPTIONcompsoc
\usepackage[caption=false,font=normalsize,labelfont=sf,textfont=sf]{subfig}
\else
\usepackage[caption=false,font=footnotesize]{subfig}
\fi
\begin{document}
\pagestyle{empty}
\title{\texttt{ChainGuard}: A Blockchain-based Authentication and Access Control Scheme for Distributed Networks}

\author{\IEEEauthorblockN{Faisal Haque Bappy$^{1}$, Joon S. Park$^{2}$, Kamrul Hasan$^{3}$, and Tariqul Islam$^{4}$}
\IEEEauthorblockA{
$^{1, 2, 4}$ Syracuse University, Syracuse, NY, USA\\
$ ^{3}$ Tennessee State University, Nashville, TN, USA\\
Email: \{fbappy@syr, jspark@syr, mhasan1@tnstate, mtislam@syr\}.edu} 
}

\maketitle

\thispagestyle{fancy}
\lhead{This work has been accepted at the 2025 IEEE Consumer Communications \& Networking Conference (CCNC 2025)}
\cfoot{}
\begin{abstract}
As blockchain technology gains traction for enhancing data security and operational efficiency, traditional centralized authentication systems remain a significant bottleneck. This paper addresses the challenge of integrating decentralized authentication and access control within distributed networks. We propose a novel solution named \texttt{ChainGuard}, a fully decentralized authentication and access control mechanism based on smart contracts. \texttt{ChainGuard} eliminates the need for a central server by leveraging blockchain technology to manage user roles and permissions dynamically. Our scheme supports user interactions across multiple organizations simultaneously, enhancing security, efficiency, and transparency. By addressing key challenges such as scalability, security, and transparency, \texttt{ChainGuard} not only bridges the gap between traditional centralized systems and blockchain's decentralized ethos but also enhances data protection and operational efficiency. 

\end{abstract}


\section{Introduction}
Private blockchains offer substantial security advantages over traditional systems, enhancing data integrity, reducing single points of failure, and improving resistance to tampering \cite{Narayanan2016}. This has led many organizations to adopt blockchain for internal operations \cite{Zheng2018}. However, authentication often remains centralized, creating security risks by concentrating access control in a single gateway \cite{thorve2022decentralized}.

This paper addresses the challenge of decentralizing authentication in distributed networks. Centralized authentication systems face issues such as single points of failure, increased attack vulnerability, and scalability limitations \cite{ferraiolo1999role}, which conflict with blockchain’s decentralized model \cite{dai2019blockchain, kimani2020blockchain}. Existing work, including decentralized identity management \cite{ren2019identity} and smart contract-based access control \cite{rahman2020context}, shows promise, but few solutions fully align with blockchain's decentralized nature \cite{mikula2018identity, ouaddah2016fairaccess}. This gap emphasizes the need for decentralized, scalable, and secure authentication solutions.

To address this challenge, we propose \texttt{ChainGuard}, a fully decentralized authentication and access control mechanism leveraging smart contracts. \texttt{ChainGuard} eliminates the need for a central server by using blockchain to manage user roles and permissions, enabling users to operate across multiple organizations without centralized oversight. Its dynamic role assignment ensures secure, efficient, and transparent access control, addressing the limitations of traditional methods. By utilizing blockchain-based identities and smart contracts, \texttt{ChainGuard} tackles key challenges such as scalability, security, and transparency. The following are the contributions of this paper.

\begin{itemize}
    \item We proposed \texttt{ChainGuard}, a novel scheme for decentralized authentication and access control in distributed networks.
 
    \item We developed a blockchain-based identity framework integrated with smart contracts to manage secure, decentralized role-based access control.
 
    \item We enhanced operational efficiency and security by introducing dynamic role assignment mechanisms within distributed networks.
\end{itemize}

\section{System Architecture} \label{model}
\texttt{ChainGuard} consists of two key components: User Role Assignments (URA) and Permission Role Assignments (PRA).

\subsection{User Role Assignments (URA)}
The user role assignment workflow starts with users registering their identities via the Smart Contract for Users (SCU), providing their blockchain account address and an encrypted password, which are stored in the Blockchain-based Identity Manager. Users then request roles through the URA, which validates credentials and role eligibility. Once approved, roles are assigned, and the Blockchain-based Identity Manager is updated. All actions are immutably recorded on the blockchain, ensuring transparency and accountability. Users can later manage roles, with all changes securely logged.

\subsection{Permission Role Assignments (PRA)}
In the permission role assignment workflow, organizations request permissions via the Smart Contract for Organizations (SCO), specifying required roles. The PRA validates these requests against user roles and organizational policies. Once approved, permissions are granted and immutably recorded on the blockchain. Organizations can then verify permissions through the PRA before granting access or executing transactions. The PRA also manages permission updates or revocations, ensuring all changes are securely logged.

\subsection{Core Components}
\textbf{Wallet:} A secure digital vault for storing blockchain identity credentials, including a blockchain account address (public key) and an encrypted password (private key) for transactions and modifications, ensuring confidentiality and integrity.

\textbf{Blockchain-based Identity Manager:} The Blockchain-based Identity Manager is a decentralized ledger maintained by trusted blockchain entities, overseeing user profiles, roles, and permissions. Its transparent, immutable design ensures identity integrity, preventing tampering and fostering trust for secure transactions.

\begin{figure*}[htbp]
\centering
\includegraphics[width=0.75\linewidth]{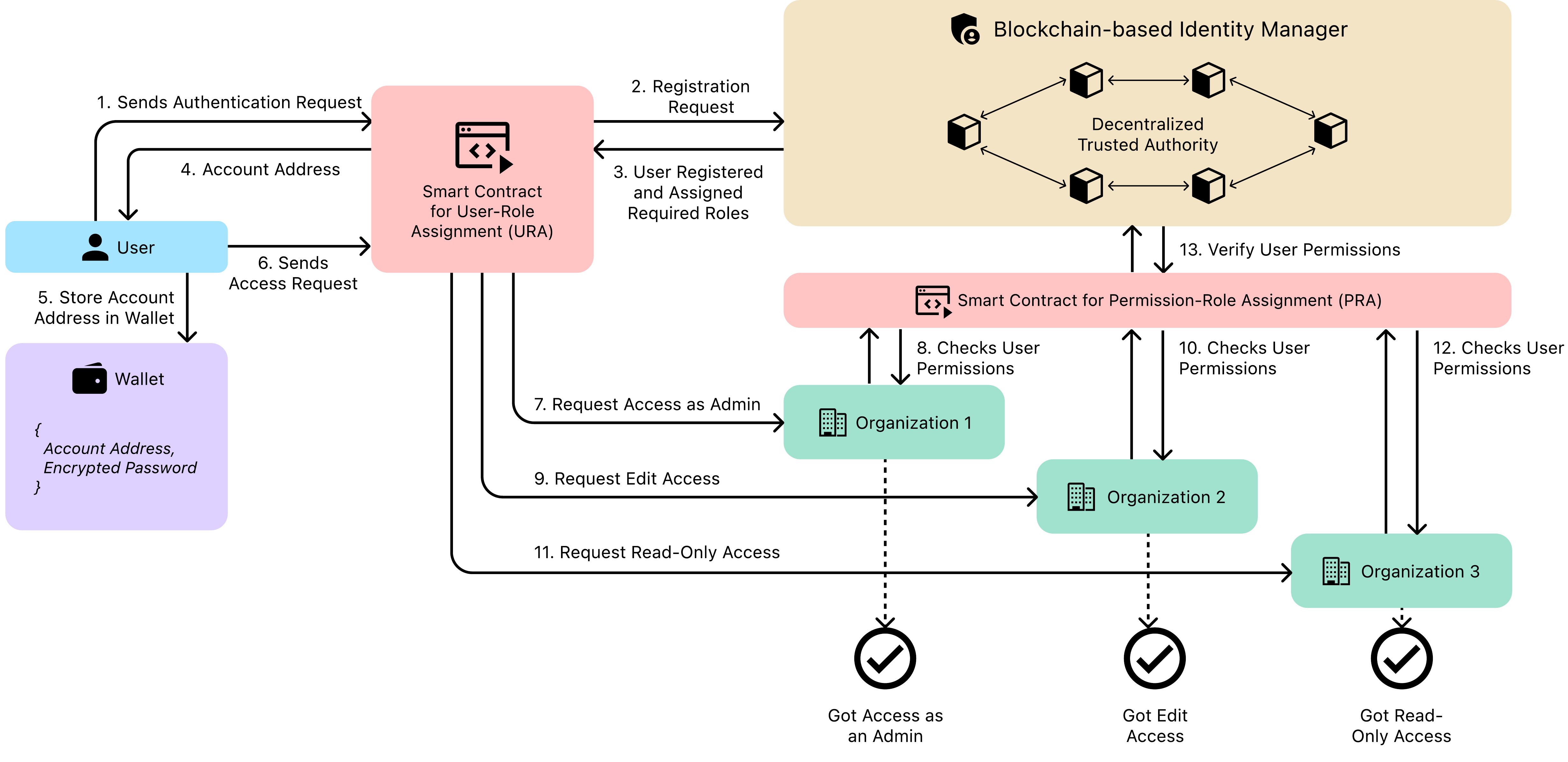}
\caption{System Workflow of URA and PRA in \texttt{ChainGuard}}
\label{Sys}
\end{figure*}

\textbf{Smart Contract for Users:} The Smart Contract for Users manages user registration and role assignments. The \textit{RegisterUser} function validates the user address, assigns a role, and emits a \textit{UserRegistered} event. The \textit{UpdateUserRole} function updates roles for registered users, emitting a \textit{UserRoleUpdated} event. These functions ensure systematic, transparent role management with all actions recorded on the blockchain.

\textbf{Smart Contract for Organizations:} The Smart Contract for Organizations manages permission roles within the blockchain ecosystem. It includes three functions: \textit{GrantPermission}, \textit{RevokePermission}, and \textit{CheckPermission}. \texttt{GrantPermission} validates the user address, assigns the role's permission, and emits a \textit{PermissionGranted} event. \textit{RevokePermission} revokes the permission and emits a \textit{PermissionRevoked} event. \textit{CheckPermission} queries the current permission status of a user’s role, enabling real-time verification.

\section{Conclusion} \label{conclusion} 
In this paper, we introduced a novel decentralized authentication and access control scheme named \texttt{ChainGuard}. Our system leverages smart contracts and blockchain-based identities to overcome the limitations of traditional centralized methods, such as single point of failure and scalability issues. By enabling dynamic role assignments and eliminating central control, our approach enhances security and transparency, mitigating risks of unauthorized access and data breaches. The scheme's adaptability to various domains, including healthcare, cloud computing, and financial services, demonstrates its versatility. In the future, we will focus on full-fledged implementation, exploring additional use cases, and implementing real-world pilots to validate the system’s effectiveness and performance.

\bibliographystyle{IEEEtran}
\bibliography{IEEEabrv,mybibfile}

\end{document}